# GNN-PT：Enhanced Prediction of Compound-protein Interactions by Integrating Protein Transformer

Jingtao Wang [1*], Xi Li[2], Hua Zhang[3]

[1] Computational Biology Department, School of Computer Science, Carnegie Mellon University, Pittsburgh, PA 15213, USA

[2] College of Computer Science, Zhejiang University, Hangzhou, Zhejiang, PR China 310027

[3] School of Computer and Information Engineering, Zhejiang Gongshang University, Hangzhou, Zhejiang, PR China 310018

**Abstract**

The prediction of protein interactions (CPIs) is crucial for the in-silico screening step in drug discovery. Recently, many end-to-end representation learning methods using deep neural networks have achieved significantly better performance than traditional machine learning algorithms. Much effort has focused on the compound representation or the information extraction from the compound-protein interaction to improve the model capability by taking the advantage of the neural attention mechanism. However, previous studies have paid little attention to representing the protein sequences, in which the long-range interactions of residue pairs are essential for characterizing the structural properties arising from the protein folding. We incorporate the self-attention mechanism into the protein representation module for CPI modeling, which aims at capturing the long-range interaction information within proteins. The proposed module concerning protein representation, called Protein Transformer, with an integration with an existing CPI model, has shown a significant improvement in the prediction performance when compared with several existing CPI models.

## 1 Introduction

In-silico screening, which generates drug candidates, is usually the first step in drug discovery. Machine learning-based methods predicting the compound-protein interactions (CPIs) have been playing an important role in this step. In the past decade, end-to-end representation learning using deep neural networks with excellent performance, which does not involve complicated feature engineering for discrete symbolic data (e.g., words of natural language and amino acids of proteins), has been widely applied to various areas, including the natural language translation[1], protein classification[2,3] as well as the CPI problem[4–7].

The data about CPIs are initially described as discrete symbols. Specially, compounds are represented as graphs where the vertices are atoms, the edges are chemical bonds, and proteins are amino acid sequences. The deep neural architectures for CPIs are in general composed of three modules, i.e., the C-module representing the compounds, the P-module encoding the proteins and the I-module for the compound-protein interactions by integrating these two modules. The deep learning models for the C-module and P-module need to be compatible with the data structure of the compound (graph) and protein (sequence) respectively. Therefore, a common choice about the network architecture is the graph neural network[8](GNN) for the C-module, and the convolutional neural networks (CNN)[9] or the recurrent neural network (RNN)[10] for the P-module. As the first end-to-end representation model for the CPI prediction, Tsubaki et al.[11] adopted GNN for the C-module and CNN for the P-module. The outputs of the C-module and the P-module are the vector representations of compounds and proteins respectively, and are then concatenated to the I-module for the CPI prediction.

Recently, many variations of attention mechanisms have been applied to capture the interactions of vector representations in various areas, including computer vision, nature language processing (NLP), bioinformatics. For example, Transformer[1] was firstly proposed as a natural language translation model, which was accomplished based on the attention mechanism (without inclusion of CNN or RNN) and achieved the state-of-the-art performance at that time. The neural attention mechanism[12] is also usually applied to the I-module in the CPI framework for predicting the interactions between the compounds and the proteins. Several recent approaches[13,14] proposed the CPI models by using the self-attention mechanism in the I-module. Molecule Transformer DTI[15], instead, utilized the self-attention in the C-module to learn the compound representation aiming at capturing the interaction information among the atoms of the compounds. They argue that the self-attention mechanism can better relate long-range atoms in chemical compounds better than other network architectures. However, the previous studies have not paid close attention to the application of the self-attention mechanism in the P-module for the CPI prediction.

In this work, we proposed a novel P-module for CPI prediction aiming at sufficiently extracting the protein features implied in the long-range residue-residue interactions. Due to the fact that CNN or RNN has no advantage in long-range information extraction, we designed the P-module by utilizing the self-attention mechanism. Besides, we observed in practice that CNN layers are still useful to extract the local features that are relevant to CPI. Therefore, the proposed novel P-module in this work includes both the self-attention layers and the CNN layers, called Protein Transformer. Finally, we integrate the existing GNN method for C-module and the proposed Protein Transformer (PT) for the P-module into an entire CPI model, named as GNN-PT. This proposed approach shows the significant performance improvement when compared with the previous work by Tsubaki et al.[11] validated on the same datasets. The experiment results also implied that the proposed Protein Transformer in the GNN-PT model is the main contribution to the performance improvement of the CPI prediction.

## 2 Materials and Methods

### 2.1 The Self-Attention Mechanism

Vaswani *et al*.[1] adopted self-attention mechanism as a substitution of RNN in their language translation model. Motivated by the finding that it is more capable of learning long-range dependencies between words, we integrated the self-attention mechanism into the P-module.

An attention function is a mapping from a Key-Value (K-V) pair and a Query (Q) to an output, where the Query, Key, Value and the output are all represented as vectors. In our case, Q, K, V are the linear projection of the same input protein sequence representations, and the output is the new protein representation incorporating the mutual association between amino acids. The whole process includes three steps: Acquiring the linear projections Query Key Value; computing the Weight by putting the Query and the Key into a compatibility function; and getting the output by computing the weighted sum of the Value using the computed Weight as the weight.

The compatibility function has many kinds of variations making there to be many versions of attentions. In this work, due to the length of the protein data (thousands of amino acids), the authors adopt the version with least time and space efficiency: "Scaled Dot-Product Attention". The compatibility function of it calculates the dot product of the Query and the Key, divides it by $\sqrt{d_k}$ where $d_k$ is the dimension of Key, and finally apply softmax on it to get the Weight.

$$Weight = soft\,max(\frac{QK^T}{\sqrt{d_k}}) \quad (1)$$

Here, Weight is a square matrix with the number of rows/columns equal to the length of the protein, i.e. the number of amino acids. The value in the $i$th row $j$th column of the Weight represents the interaction intensiveness between the $i$th and $j$th amino acids.

After calculating the weight, each row of the output, which is an amino acid vector, can be calculated as the weighted sum of all amino acids. This is accomplished by a single matrix multiplication:

$$Output = Weight \times V = soft\,max(\frac{QK^T}{\sqrt{d_k}})V \quad (2)$$

The intuition behind this is that amino acids should be allowed to interact based on the strength of the mutual interactions.

## 2.2 The overall architecture of the proposed CPI model

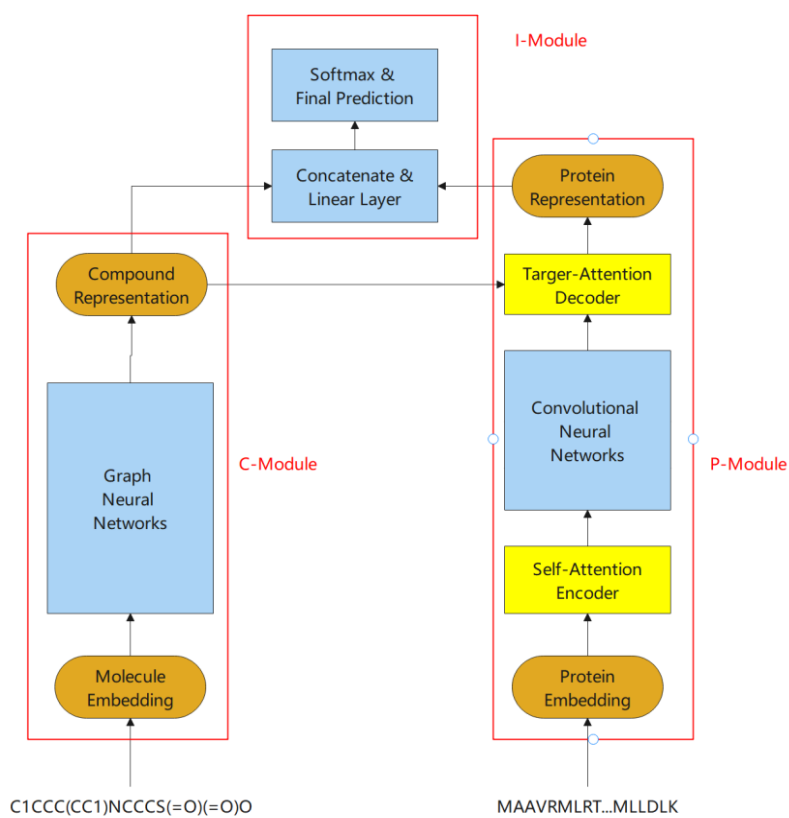

Figure 1. The overall architecture of the proposed CPI model (GNN-PT) composed of three modules that are C-module, P-module and I-module, respectively.

Figure 1 shows the overall network architecture of the proposed model. The left part of the picture shows how the model handle compound representation learning, which is the C-Module. Molecules in SMILES format are converted to graph, embedded, and fed into the C-Module. The C-Module takes the embedding as input and learn the compound representation vector with 3 layers of GNNs. This compound representation will not only be feed into the I-module for final prediction, but also be used in the P-Module for guiding protein representation learning. On the right is the P-Module for learning the protein representation, which is where we add self-attention mechanism for learning long-distance interactions. The protein sequences are split into n-grams, embedded, adding the positional encoding followed by Vaswani *et al*.[1], and fed into the P-Module for learning the protein representation. Finally, on top of the C-module and P-module, I-module takes the output from both and produces the final interaction prediction.

In this work, we performed controlled experiments to demonstrate that the improvement in protein representation learning indeed comes from our re-designed P-module, which utilizes self-attention. To do this, we limit our modification to the P-module CPI model proposed by Tsubaki et al. only, while using the C-module and the I-module as control. Therefore, in the rest of this section, we will illustrate in detail how we design the P-Module and enable it to better learn the protein representation.

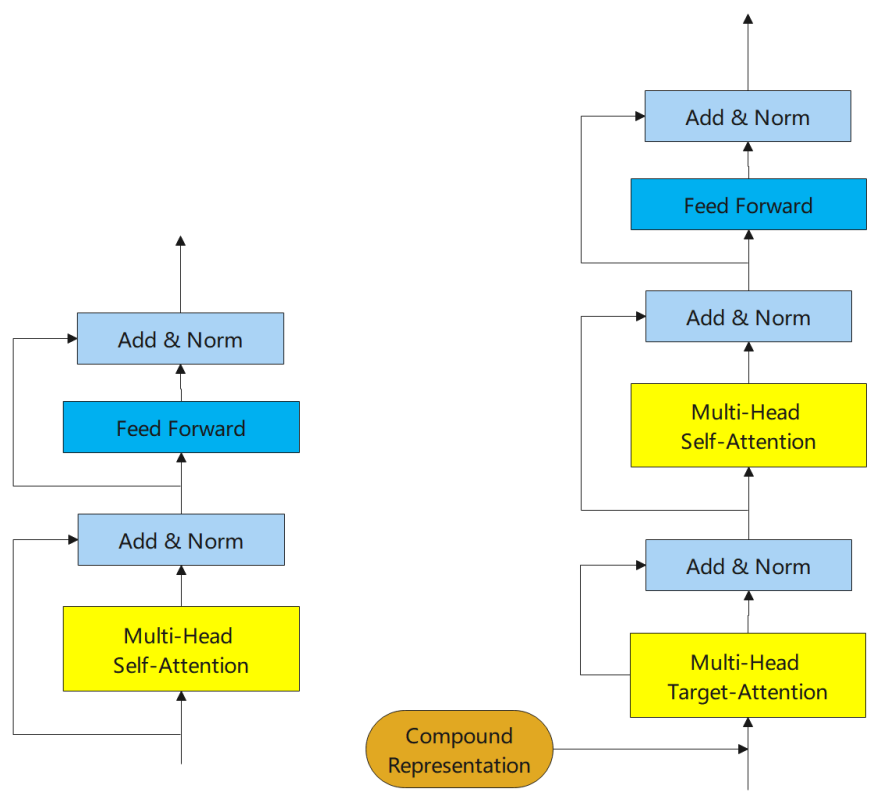

Figure 2. The network architectures of the self-attention encoder (left) and the target-attention decoder (right) in the P-module.

## 2.3 The Protein Transformer for the P-module

The proposed P-module, called Protein Transformer (PT), as shown in Figure 2, includes three sequential layers: self-attention encoder, CNN layer and target-attention decoder. The embedded protein representation is first fed into N layers of Self-Attention Encoder, the key component of our model. In this encoder, Self-Attention helps to learn the mutual interaction between all amino acid pairs, disregarding the distance between them. Intuitively speaking, this corresponds to the folding process of proteins.

$$AttentionScore = Weight = Softmax(\frac{Query \cdot Key^T}{\sqrt{d_k}})$$

To compute the attention scores for the self-attention mechanism, the query and the key are both protein representation.

Coming out of the self-attention encoding layer, neurons in the protein representation contain "self-interaction" information, including the long-range residue interactions. It is then fed into the CNN for further learning the properties.

The final part of the P-module is the target-attention decoder, where the interaction between amino acid residues and the compound is learned. Compound representation and protein representation are both fed into the multi-head target-attention layer. For computing the target-attention, the query is the compound representation, and the key is the protein representation. The attention scores of this target-attention mechanism reflect the intensity of the interaction between the compound and the residues of the protein.

The target attention decoder outputs informative protein representation with properties regarding the structure, binding site, included. It will then be concatenated with the compound representation and fed into the I-module for final prediction.

## 3 Results

### 3.1 Performance comparison with other existing methods

We performed the proposed model on 2 datasets (*human & C.elegans*) that were created by Liu *et al*. [16] and also used by Tsubaki et al.[11]. We used all the same hyperparameters for the datasets demonstrating that our model is universal for all kinds of CPI data. We used number of encoding layer = 3, number of decoding layer = 1, learning rate = 2e-4, warmup step for the attention mechanism = 50 for all datasets. The performance of the proposed GNN-PT, is recorded in the last column as shown in Table 1, Table 2, Table 3 and Table 4.

**Table 1.** Performance comparison for the case of Human positive/negative ratio 1:1.

| Measure | KNN | RF | logR | SVM | Tsubaki et al. | GNN-PT |
| --- | --- | --- | --- | --- | --- | --- |
| AUC | 0.860 | 0.940 | 0.911 | 0.910 | 0.970 | **0.978** |
| Precision | 0.798 | 0.861 | 0.891 | **0.966** | 0.923 | 0.928 |
| Recall | 0.927 | 0.897 | 0.913 | 0.950 | 0.918 | **0.952** |

**Table 2.** Performance comparison for the case of Human positive/negative ratio 1:3.

| Measure | KNN | RF | logR | SVM | Tsubaki et al. | GNN-PT |
| --- | --- | --- | --- | --- | --- | --- |
| AUC | 0.904 | 0.954 | 0.920 | 0.942 | 0.950 | **0.980** |
| Precision | 0.716 | 0.847 | 0.837 | **0.969** | 0.949 | 0.917 |
| Recall | 0.882 | 0.824 | 0.773 | 0.883 | 0.913 | **0.925** |

**Table 3.** Performance comparison for the case of C.elegans positive/negative ratio 1:1.

| Measure | KNN | RF | logR | SVM | Tsubaki et al. | GNN-PT |
| --- | --- | --- | --- | --- | --- | --- |
| AUC | 0.858 | 0.902 | 0.892 | 0.894 | 0.978 | **0.986** |
| Precision | 0.801 | 0.821 | 0.890 | 0.785 | 0.938 | **0.970** |
| Recall | 0.827 | 0.844 | 0.877 | 0.818 | 0.929 | **0.960** |

Table 4. Performance comparison for the case of C.elegans positive/negative ratio 1:3.

| Measure | KNN | RF | logR | SVM | Tsubaki et al. | GNN-PT |
|---|---|---|---|---|---|---|
| AUC | 0.892 | 0.926 | 0.896 | 0.901 | 0.971 | **0.983** |
| Precision | 0.787 | 0.836 | 0.875 | 0.837 | 0.916 | **0.942** |
| Recall | 0.743 | 0.705 | 0.681 | 0.576 | 0.921 | **0.929** |

The results of K-nearest neighbors (KNN), random forest (RF), logistic regression (logR), support vector machine (SVM), have been already reported in the work by Liu *et al*.[16]. As shown above, the proposed GNN-PT outperforms the existing methods on almost all conditions in terms of metrics and datasets. Notably, our model has a significantly higher AUC in all cases, demonstrating this model that learn the knowledge very well and the prediction results are always aligning with the truth.

## 4 Discussion

Notice that the main contribution of our work is to propose this approach by incorporating the self-attention mechanism to the P-module to help the model learn better. The proposed GNN-PT method can almost be applied to any model for protein sequences. Based on this idea, and also because the authors are not available to much computational resources, we did not try to fine-tune the model to get the optimal results. Our argument has been justified since the current one has a significant improvement. That said, we discovered phenomena like stacking more self-attention encoder tend to further improve the performance, etc. Those are worth trying so that a better model can be discovered.